\documentclass[]{aa}  

\usepackage{graphicx}
\usepackage{txfonts}
\usepackage{ulem}
\begin{document}


\title{Survival of fossil fields during the pre-main sequence evolution \\ of intermediate-mass stars}


\author{Dominik R.G. Schleicher
          \inst{1}
          \and
          Juan Pablo Hidalgo\inst{1}
\and
Daniele Galli\inst{2}
             }

   \institute{Departamento de Astronom\'ia, Facultad Ciencias F\'isicas y Matem\'aticas, Universidad de Concepci\'on, Av. Esteban Iturra s/n,Barrio Universitario, Concepci\'on, Chile.
              \email{dschleicher@astro-udec.cl}
         \and
         INAF - Osservatorio Astrofisico di Arcetri, Largo E. Fermi 5, 50125 Firenze, Italy
          }


 
  \abstract
   {Chemically peculiar Ap and Bp stars host strong large-scale magnetic fields in the range of $200$~G up to $30$~kG, which are often considered to be the origin of fossil magnetic fields.}
   {We assess the evolution of such fossil fields during the star formation process and the pre-main sequence evolution of intermediate stars, considering fully convective models, models including a transition to a radiative protostar and models with a radiative core. We also examine the implications of the interaction between the fossil field and the core dynamo.
   }
   {We employ analytic and semi-analytic calculations combined with current observational constraints.}
   {For fully convective models, we show that magnetic field decay via convection can be expected to be very efficient for realistic parameters of turbulent resistivities. Based on the observed magnetic field strength - density relation, as well as the expected amount of flux loss due to ambipolar diffusion, it appears unlikely that convection could be suppressed via strong enough magnetic fields. On the other hand, a transition from a convective to a radiative core could very naturally explain the survival of a significant amount of flux, along with the presence of a critical mass. We show that in some cases, the interaction of a fossil field with a core dynamo may further lead to changes in the surface magnetic field structure.}
   {In the future, it will be important to understand in more detail how the accretion rate evolves as a function of time during the formation of intermediate-mass protostars, including its impact on the protostellar structure. The latter may even allow to derive quantitative predictions concerning the expected population of large scale magnetic fields in radiative stars.}


   \maketitle
%

\section{Introduction}

It is known since the calculations of \citet{Cowling1945} that magnetic fields can survive for resistive timescales comparable to or even longer than the main-sequence life time of radiative stars. Indeed, strong magnetic fields have been detected in a population of peculiar intermediate-mass main-sequence stars in the range of $1.5-6$~M$_\odot$. These stars, that are generally classified as Ap/Bp, host large-scale magnetic fields with mean field strength in the range of $200$~G up to $30$~kG \citep{Auriere2007}. In general, these stars account for a few percent of the A-star population. Studies based on a volume limited sample within $100$~pc from the Sun further indicate a significant mass dependence, where the fraction of magnetically active stars is at the sub-percent level at $1.5$~M$_\odot$, and raises to $\gtrsim20\%$ and higher from stellar masses of $3.5$~M$_\odot$. Below $1.5$~M$_\odot$ (around F0), the phenomenon disappears completely and is then absent in stars with efficient convection within the envelope. 

The presence of strong magnetic fields has been shown to be correlated with other properties. \citet{Abt1995} have shown that most magnetic A stars rotate slowly compared to the non-magnetic stars. More specifically, \citet{Mathys2008} have shown that most Ap stars have periods between one and ten days, while non-magnetic A stars have periods ranging from a few hours to a day. $10\%$ of Ap stars even have very long periods of the order $100$~days. Similarly, the binary fraction among Ap stars was found to be lower than for conventional A stars \citep{Abt1973, Gerbaldi1985, Carrier2002, Folsom2013}. Apart from one known example (HD 200405), there seems to be a lack of Ap stars in binaries with periods of less than 3 days.

The strong magnetic fields are further linked to chemically peculiar abundance patterns, which may arise as a result of gravitational settling and radiative levitation \citep{Michaud1970, Kochukhov2011}. Generally, Ap/Bp stars are defined as showing peculiar abundances of rare earths and some lighter elements such as silicon, as well as surface inhomogeneities of these elements. There is a strong correlation between the chemically peculiarities and the presence of magnetic fields, with the apparent exception of HgMn stars. 

It seems very likely that these are not separate phenomena, but the chemical peculiarities and the strong magnetic fields may be linked, particularly if the magnetic field is indeed a fossil field coming from the interstellar medium. In this case, it will have to survive the convective protostellar phase, which is only possible for strong enough magnetic fields that can efficiently suppress convection \citep[see e.g.][]{Moss2003, Jermyn2021}. An alternative possibility is available particularly in the range of higher mass stars. As noted by \citet{Dudorov1990}, protostars may become radiative when reaching masses of $\sim2$~M$_\odot$, allowing them to conserve any accreted magnetic field. Protostellar evolution calculations for intermediate-mass stars have indeed shown such a critical transition to occur \citep{Palla1991}, with the mass of the transition being regulated by the accretion rate of the protostars \citep{Palla1992, Palla1993}. These possibilities are likely to be directly linked to chemical peculiarities of the stars, as efficient convection would not only wash out any fossil fields, but also any gentle separation processes of the elements in the surface of the stars \citep[see also][for the suppression of convection via strong magnetic fields]{Gough1966, Moss1969}. {Numerical simulations suggest that kG magnetic fields may be implanted in low mass stars as a result of gravitational collapse, though considerable uncertainties remain due to effects of numerical resolution \citep{Wurster2018, Wurster2022}.}

Meanwhile, weak magnetic fields have even been detected in part of the ``non-magnetic'' population, namely in Sirius and Vega. In case of Vega, Zeeman polarimetry has revealed a magnetic field strength of $0.6\pm0.3$~G \citep{Lignieres2009, Petit2010, Petit2022}, and $0.2\pm0.1$~G in Sirius \citep{Petit2011}. It is conceivable but requires further investigation that all of the so-called ``non-magnetic'' stars have field strength of this order, corresponding then to a gap of about two orders of magnitude to the Ap stars showing strong magnetic fields. It is also important to note that the detection of magnetic fields via the Zeeman effect becomes considerably more difficult for stars above $6-8$~M$_\odot$, as they show much fewer lines in the spectrum. In some cases, detection was achieved \citep{Henrichs2000, Wade2013}, and about $7\%$ of O and B stars seem to host large-scale magnetic fields \citep{Petit2013}.

An important alternative to the fossil field scenario could be due to the presence of a core dynamo in intermediate mass stars, which are expected to develop a convective core when reaching the main sequence \citep{Palla1991, Palla1992}. The cores of such stars have long been suspected to host dynamo action \citep{Krause1976, Brun2005}. The buoyant rise of the produced magnetic structures is however expected to take longer than the lifetime of the star, unless very small flux tubes can be formed \citep{Parker1979, Moss1989, MacGregor2003, MacDonald2004}, while observations tend to favor large-scale magnetic fields at the surface. Nonetheless, it is worth noting that the fossil field can even interact with the core dynamo and possible enhance it \citep{Featherstone2009}, and it is likely important to further explore its implications. 

It is worth noting that additional complications and problems may exist, as lined out in the review article of \citet{Braithwaite2017}. One such issue concerns the stability of fossil fields, which was examined for toroidal fields by \citet{Spruit1999} and \citet{Braithwaite2006}, and for axisymmetric poloidal fields by \citet{Markey1973}, \citet{Wright1973}, and \citet{Markey1974}. Numerical investigations have been performed by \citet{Braithwaite04} and \citet{Braithwaite2009}, following the evolution of magnetic fields in radiative stars from initially arbitrary fields based on the evolution equations. These simulations have provided more detailed conditions for the stability of magnetic fields, which are compatible for a wide range of cases and field strength with the analytic calculations. 

\begin{figure}
    \centering
    \includegraphics[scale=0.17]{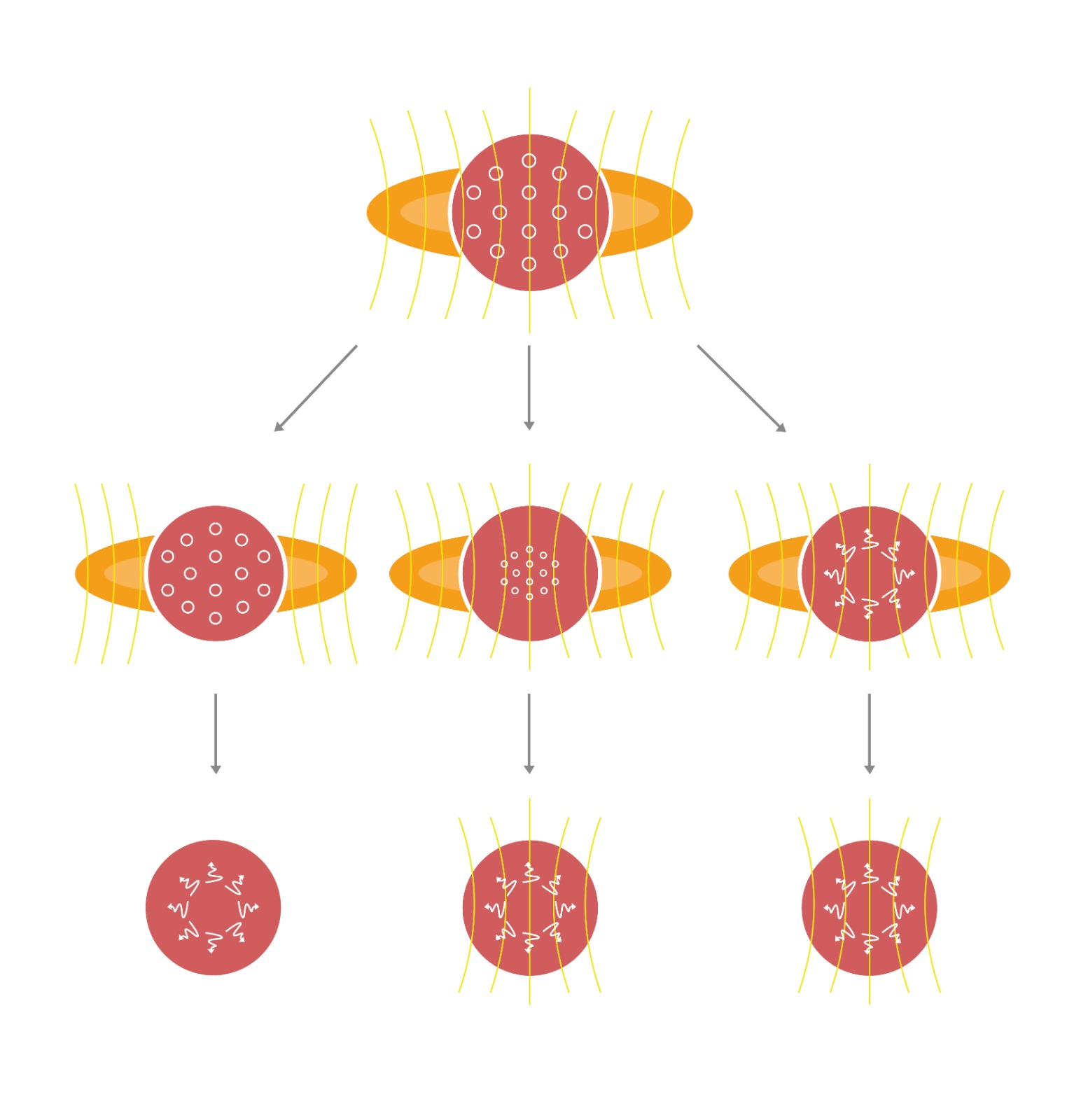}
    \caption{Possible evolutionary pathways of fossil fields during protostellar evolution. Left: Fossil field dissipated due to convection. Middle: Fossil field strong enough to suppress convection. Right: Protostar becomes radiative, absence of turbulent dissipation. The orange color indicates the protostellar disk, yellow lines the magnetic field. {Circles indicate convective regions within the protostars, arrows represent radiative regions.}}
    \label{fig:caminos}
\end{figure}

In this paper, we are interested in following the evolution of the initial fossil field through the star formation process \citep[e.g.][]{Galli2006, Shu2007, Hennebelle2016}, particularly in the context of massive star formation. Possible implications of non-ideal magneto-hydrodynamics (MHD) were previously assessed e.g. by \citet{Desch2001}, \citet{Nakano2002}, \citet{Tomida2015} and \citet{Masson2016}.  A theory for the evolution of the fossil fields during the collapse and disk phase was presented by \citet{Dudorov2014, Dudorov2015}. The evolution of the magnetic field within a fully convective protostar was then discussed by \citet{Moss2003}.  However, as already suggested by \citet{Dudorov1990} and confirmed by protostellar evolution models of \citet{Palla1991, Palla1992, Palla1993}, the protostars will eventually become radiative depending on the accretion rate. In some cases also simulations where the protostars develop a radiative core have been reported \citep{Klassen2012}. A summary of the most important evolutionary pathways is given in Fig.~\ref{fig:caminos}.

Here we reassess the evolution of the fossil field in the light of such possibilities, considering the approximate evolution during the star formation process including the range of magnetic field strength in the context of different types of protostellar models. Our considerations for the prestellar evolution are presented in Section~\ref{prestellar}. The evolution in case of a fully convective protostar is presented in Section~\ref{fully} and compared to \citet{Moss2003}. The scenario considering a transition to a radiative protostar is outlined in Section~\ref{radiative}, and the case with a radiative core presented in Section~\ref{core}. The possible interaction of the fossil field with a core dynamo is considered in Section~\ref{dynamo}, and a final discussion and conclusions are provided in Section~\ref{conclusions}.

\section{Pre-stellar evolution}
\label{prestellar}

Magnetic field measurements in molecular and protostellar clouds have established an approximate relation between the magnetic field strength $B$ and the number density of the gas $n$, 
\begin{equation}
\frac{B}{B_0}=\left( \frac{n}{n_0} \right)^k, \label{Bn}
\end{equation}
where $B_0=5\ \mu$G, $n_0=50-200$~cm$^{-3}$ and $k$ between $1/2$ and $1/3$. This relation has been obtained for the density range of roughly $1-10^{10}$~cm$^{-3}$ \citep{Vallee1997, Girart2006, Li2009, Crutcher12, Chapman2013, Koley2022}. It is very similar though not completely identical to what could be expected from flux freezing during a spherically symmetric collapse, where one would 
expect $B\propto n^{2/3}$. The fact that the observed $B$-$n$ relation has a somewhat more moderate slope could be due to different reasons; for example if the collapse is not spherical the relation between density and radius may change, the expression for flux conservation could be altered and flux may not be fully conserved. Numerical simulations have studied the evolution of the magnetic field strength during gravitational collapse, starting with typical 
values of $30-300 ~\mu$G at a number density of $300$~cm$^{-3}$, finding a central field strength of about $0.1$~G at a number density of around $10^{10}-10^{11}$~cm$^{-3}$ \citep{Desch2001, Nakano2002, Tomida2015, Masson2016}. At those densities, the ionization degree drops significantly and the magnetic field is no longer efficiently coupled to the gas, but remains approximately constant while the gas density increases to about $10^{15}-10^{16}$~cm$^{-3}$. At that point, the temperature has risen so much due to adiabatic collapse that the ionization degree is sufficiently high for the coupling of the magnetic field to the gas \citep{Nakano1988, NakanoUmebayashi1988} . We assume that the magnetic field strength will follow a similar relation as given in Eq.~(\ref{Bn}) until protostellar densities are reached. These final densities will depend on the protostellar mass and radius and also on the protostellar model \citep[see e.g.][]{Palla1993,Siess2000}.

A possible complication arises in the context of protoplanetary disk formation. In a detailed study, \citet{Hennebelle2016} have shown that disk formation occurs when the timescale for the generation of a toroidal field component through differential rotation is comparable to the ambipolar diffusion time in the vertical direction, thereby reducing the efficiency of magnetic braking. In addition, the rotation and magnetic braking timescales must be of the same order. This leads to the condition
\begin{equation}
r_{\rm disk}\sim 19\ \delta^{2/9}\left( \frac{\eta_{\rm AD}}{10^{18}\ \mbox{cm$^2$~s$^{-1}$}} \right)^{2/9}\left(\frac{B_z}{0.1\ \mathrm{G}} \right)^{-4/9}\left( \frac{M_*+M_{\rm d}}{0.1\ M_\odot} \right)^{1/3}~\mathrm{au},
\end{equation}
where $\delta$ is a coefficient of the order of a few in the density-radius relation, $\eta_{\rm AD}$ is the ambipolar diffusion resistivity, $B_z$ the $z$-component of the magnetic field (which follows a similar relation as our Eq.~(\ref{Bn}) in the \citet{Hennebelle2016} model and then similarly flattens in the regime of ambipolar diffusion), $M_*$ is the mass of the central star and $M_{\rm d}$ the mass of the disk. We estimate the density at the disk radius assuming that the density profile until that point approximately follows an isothermal sphere, i.e.
\begin{equation}
\rho\approx \delta\frac{c_s^2}{2\pi Gr^2},
\end{equation}
with the sound speed $c_s=\sqrt{\gamma k_{\rm B} T/(\mu m_{\rm H})}$, $k_{\rm B}$ the Boltzmann constant, $T$ the temperature, $\mu\sim2$ the mean molecular weight, $m_{\rm H}$ the atomic hydrogen mass and $\gamma=5/3$ the adiabatic index. With $T\sim10$~K, we thus have $c_s\sim0.3$~km~s$^{-1}$ and $\rho\sim2.9\times10^{-14}$~g~cm$^{-3}$ or $n\sim8.8\times10^9$~cm$^{-3}$.  Disk formation is thus typically expected at similar, perhaps slightly lower densities compared to those where ambipolar diffusion is expected to become efficient.

The disk phase could in principle change the relation between magnetic field strength and density, particularly if a dynamo is operational at least in some part of it. The expected magnetic field strength would then be related to the field strength of equipartition,
\begin{equation}
B_{\rm eq}=\varv_{\rm turb}\sqrt{4\pi \rho},\label{Bequi}
\end{equation}
where $\varv_{\rm turb}$ is the turbulent velocity. Assuming $\varv_{\rm turb}\sim c_s$, we would have $B_{\rm eq}\sim0.018$~G at $n\sim10^{10}$~cm$^{-3}$. while from Eq.~(\ref{Bn}) we would have $B\sim0.002-0.05$~G depending on the value of $k$. 

In a more accurate model \citep[e.g.][]{Shukurov2004}, the field strength produced by the dynamo is given as
\begin{equation}
B_{\rm eq,disk}=B_{\rm eq}\sqrt{\frac{D}{D_{\rm crit}}-1},
\end{equation}
where 
\begin{equation}
D\simeq 9\left( \frac{\Omega R}{\varv_{\rm turb}} \right)^2\label{dynamonumber}
\end{equation}
is the dynamo number, which must exceed a critical value $D_{\rm crit}\sim7$ for the dynamo to operate. We estimate the angular velocity $\Omega$ via\begin{equation}
\Omega\sim\sqrt{\frac{G(M_*+M_{\rm d})}{r^3}}
\end{equation}
and obtain $\Omega\sim7.6\times10^{-10}$~s$^{-1}$ at a radius of about $19$~au. The expected dynamo number is thus $D\sim420 $, leading to a relation of $B_{\rm eq,disk}\sim 7.7 B_{\rm eq}\sim 0.14$~G. The field strengths that can be expected in the presence of a disk dynamo are thus somewhat around the upper limit of what can be expected from  the scaling relations (Eq.~\ref{Bn}). Considering the disk model presented by \citet{Dudorov2014}, the ionization degree on these scales is very low and keeps decreasing down to a scale of $0.3$~au, where the increase of the temperature leads to a higher ionization degree. The inner part of the disk can thus be expected to be exposed to a high ambipolar diffusivity and a decoupling of the gas and the magnetic field, so that no very significant change of the magnetic field strength is expected between $3$ and $0.3$~au. This behavior in principle is analogous to the behavior during gravitational collapse in the regime where the magnetic field decouples from the gas. While the presence of a disk phase will give rise to further complexities, at least the order of magnitude of the magnetic field appears to nonetheless stay in the range that might be expected from simpler considerations.

\section{Magnetic field evolution in fully convective protostars}\label{fully}

\subsection{Field strength at protostar formation}

As a starting point, we employ the fully convective protostar models adopted by \citet{Moss2003}, and we use the $B$-$n$ relation (Eq.~\ref{Bn}) to estimate the expected magnetic field strength for the densities until densities of $10^{10}$~cm$^{-3}$. Subsequently we assume the magnetic field to be decoupled due to ambipolar diffusion and thus constant, until recoupling  in the density range of $10^{14}-10^{17}$~cm$^{-3}$ \citep{Nakano1988}. After recoupling, we assume an approximately adiabatic spherically symmetric contraction  of the protostellar cloud implying $B\propto n^{2/3}$. For the specific case where a recoupling density of $10^{14}$~cm$^{-3}$ is assumed, the expected field strength for the average protostellar density is given in Table~\ref{tabla1} for $k=1/2$ and $k=1/3$. In the first case, we obtain field strengths in the range of $40$--$9000$~G, while in the second case the expected field strength is more moderate, and in the range of $2-400$~G.

\begin{table}[t!]
\centering
\caption{Estimations with $n_0 = 100~\mathrm{cm}^{-3}$, $B_1$ and $B_2$ are the mean magnetic field strength using $k=1/2$ and $k=1/3$, respectively, assuming an additional flux loss $\alpha_{\rm ad}=10^{-2}$ due to ambipolar diffusion. $B_{\rm eq}$ denotes that equipartition field strength within the protostar that is required to suppress convection. Here the recoupling between the density and the magnetic field is assumed to happen at a number density of $10^{15}$~cm$^{-3}$, and subsequently to follow a spherically symmetric contraction. For the purpose of comparison, we adopt here the fully convective protostellar models employed by \citet{Moss2003}. }
\begin{tabular}{lccccc}
\hline\hline\noalign{\smallskip}
$M_*$ & $R_*$ & $\bar{\rho}$  &$B_1$ & $B_2$ & $B_{\rm eq}$
\\
$(M_\odot)$ & $(R_\odot)$ & $(\mathrm{kg~m^{-3}})$  & $(\mathrm{G})$ & $(\mathrm{G})$ & $(\mathrm{G})$
\\
\hline
1  &  13  &   0.64  &  166.34  &  7.72 & 341  \\
1  &  6  &   6.53  &  780.85  &  36.24 & 955  \\
1  &  3  &   52.23  &  $3.1\times 10^3$  &  144.98 & 2.4$\times 10^3$  \\
1  &  2  &   176.27  &  $7\times 10^3$  &  326.20 & 4.1$\times 10^3$  \\
1  &  1.8  &   241.80  &  $8.7\times 10^3$  &  402.71 & 4.8$\times 10^3$  \\
\hline
3  &  22  &   0.40  &  120.81  &  5.61 & 1.5$\times 10^3$  \\
3  &  15  &   1.25  &  259.88  &  12.06 & 2.5$\times 10^3$  \\
3  &  13  &   1.93  &  345.99  &  16.06 & 3.1$\times 10^3$  \\
3  &  11  &   3.18  &  483.24  &  22.43 & 3.8 $\times 10^3$ \\
3  &  9  &   5.80  &  721.88  &  33.51 & 5 $\times 10^3$ \\
\hline
5  &  45  &   0.08  &  40.59  &  1.88 & 1.6$\times 10^3$  \\
5  &  30.5  &   0.25  &  88.36  &  4.10 & 2.7$\times 10^3$  \\
5  &  21.6  &   0.70  &  176.17  &  8.18 & 4.3$\times 10^3$  \\
5  &  19.3  &   0.98  &  220.67  &  10.24 & 5$\times 10^3$  \\
5  &  18  &   1.21  &  253.69  &  11.78& 5.5$\times 10^3$ \\
\hline
\end{tabular}
\label{tabla1}
\end{table}

\subsection{Required field strength to suppress convection}\label{convection}

\begin{figure}
    \centering
    \includegraphics[scale=0.55]{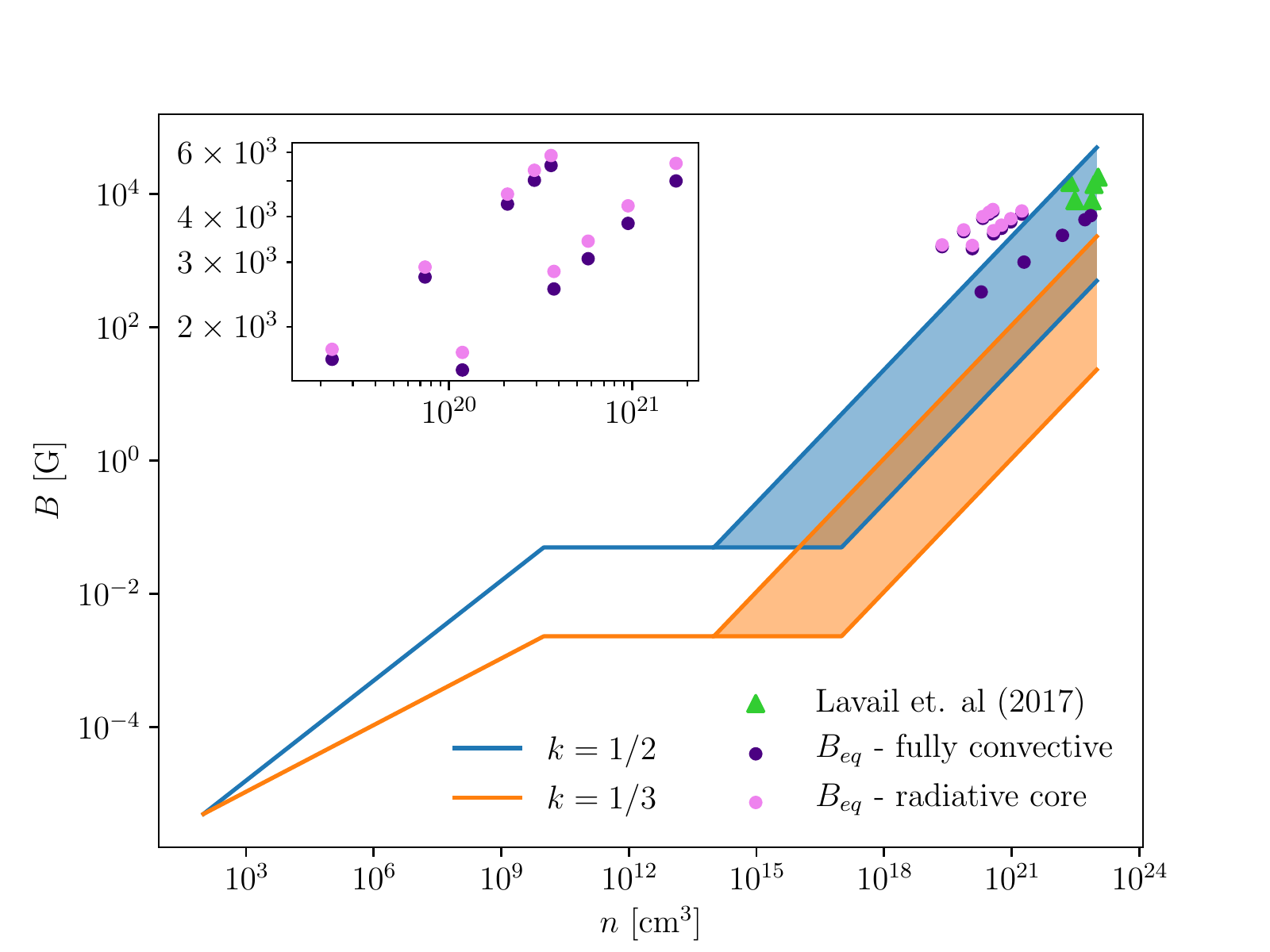}
    \caption{Magnetic field strength as a function of the number density. The blue and orange line indicate the magnetic field strength - density relation expected from gravitational collapse for the cases of $k=1/2$ and $k=1/3$ based on Eq.~(\ref{Bn}). Ambipolar diffusion is assumed to set in at a  density of $10^{10}$~cm$^{-3}$, while recoupling is assumed to occur at densities of $10^{14}-10^{17}$~cm$^{-3}$ following \citet{Nakano1988}. We subsequently assume an approximately spherical contraction with $B\propto n^{2/3}$. The indigo dots show the equipartition field strength for fully convective models, the violet dots for models with a radiative core. Green triangles are surface magnetic field strength observed in intermediate-mass T Tauri stars from the sample of \citet{Lavail2017}.}
    \label{fig:bn}
\end{figure}

We now aim to assess whether the magnetic field that is present when the protostar forms would be sufficient to suppress protostellar convection. For this purpose it is central to estimate the convective velocity $\varv_c$. Assuming convective energy transport, we have \citep{Kippenhahn2013, Schleicher2017} \begin{equation}
\varv_c=\varv_s\sqrt{\nabla-\nabla_{\rm ad}},\label{vc}
\end{equation}
with $\varv_s$ the speed of sound in the protostar, $\nabla=d\ln T/d\ln P$ the physical temperature gradient and $\nabla_{\rm ad}=(d\ln T/d\ln P)_{\rm ad}$ the temperature gradient under adiabatic conditions. 
 The sound speed in the interior can be evaluated as
\begin{equation}
\varv_s=\sqrt{\frac{GM}{R}}.\label{vs}
\end{equation}
The difference $\nabla-\nabla_{\rm ad}$ can be estimated as follows. From mixing length theory, we can write \citep{Kippenhahn2013, Schleicher2017} \begin{equation}
F_{\rm conv}=\rho C_p T\left( \frac{l_m}{H_p} \right)^2\sqrt{\frac{1}{2}gH_p}(\nabla-\nabla_{\rm ad})^{3/2},\label{Fconv}
\end{equation}
with $\rho$ the density, $F_{\rm conv}$ the convective energy flux, $C_p$ the heat capacity at constant pressure, $T$ the temperature, $l_m$ the mixing length and $g$ the gravitational acceleration. 
The quantities in Eq.~(\ref{Fconv}) can be expressed through the main properties of the protostar: mass $M_*$, radius $R_*$, and luminosity $L_*$. We have
\begin{eqnarray}
F_{\rm conv}&\sim& \frac{L_*}{4\pi R_*^2},\quad  \rho\sim\frac{3M_*}{4\pi R_*^3}, \quad T\sim\frac{\mu m_p}{k_{\rm B}}\frac{GM_*}{R_*},\\
C_p & \sim &\frac{5}{2}\frac{k_{\rm B}}{\mu m_p}, \quad \sqrt{g H_p} =\sqrt{\frac{k_{\rm B}T}{\mu m_p}}
\sim\sqrt{\frac{GM_*}{R_*}},\label{two}\\
 g&\sim&\frac{GM_*}{R_*^2},\label{three}
\end{eqnarray}
with $\mu$ the mean molecular weight. Inserting these expressions into Eq.~(\ref{Fconv}) 
and solving for $\nabla-\nabla_{\rm ad}$, we obtain
\begin{equation}
\nabla-\nabla_{\rm ad}=\left( \frac{2\sqrt{2}}{15} \right)^{2/3}\left( \frac{H_p}{l_m} \right)^{4/3}\frac{L_*^{2/3}R_*^{5/3}}{GM_*^{5/3}}.\label{nabla}
\end{equation}
Eq.~(\ref{two}) and (\ref{three}) show that $H_p\sim R_*$. While this may be a rather crude approximation, it results from the assumption of considering only average properties within the protostar.
The mixing length is assumed to be of the order of the pressure scale height,  $l_m\sim H_p\sim R_*$. 
For the protostellar luminosity, we assume the maximum stellar luminosity derived assuming Kramer's opacity $\kappa\propto\rho T^{-3/5}$, implying \citep{Hayashi1962}
\begin{equation}
L_{\rm max}\sim 0.6 \left( \frac{M_*}{M_\odot} \right)^{11/2}\left( \frac{R_*}{R_\odot} \right)^{-1/2} L_\odot.
\end{equation}
For comparison, the equipartition field strength $B_{\rm eq}$ calculated from the convective velocity in Eq.~(\ref{vc}) is plotted in Fig.~\ref{fig:bn} together with the expected $B$-$n$ relation. {When the convective velocity estimated via Eq.~(\ref{vc}) is less than the Alfv\'en velocity,}
\begin{equation} 
v_A=\frac{B}{\sqrt{4\pi \rho}},
\end{equation}
{we assume convection to be suppressed, and, as a result, the magnetic field not subject to turbulent decay. We assume the $B$-$n$ relation given by
Eq.~(\ref{Bn}) at densities up to $10^{10}$~cm$^{-3}$}, with $k=1/2$ and $k=1/3$, while we adopt an approximately constant magnetic field strength {at larger densities, assuming} recoupling between the density and the magnetic field to occur in the density range of $10^{14}-10^{17}$~cm$^{-3}$ \citep{Nakano1988}. After recoupling, we assume a rather spherical contraction of the protostellar cloud implying $B\propto n^{2/3}$. For comparison we further added the observed magnetic field strengths from a sample of intermediate-mass T Tauri stars \citep{Lavail2017}.

The results both for the equipartition magnetic field strength and the expected values based on the field strength - density relation  are given in Table~\ref{tabla1} for the protostellar models employed by \citet{Moss2003}, which were evaluated considering the average protostellar density in the model. The required field strength are in the range of $300-5.5\times10^3$~G, while the expected field strength lie in the range of $2-400$~G in case of $k=1/3$ and $40-9\times10^3$~G for $k=1/2$, in both cases assuming a relatively early recoupling at $10^{15}$~cm$^{-3}$. At least in the $1$~M$_\odot$ models during the more contracted stages, the expected magnetic field strength can exceed the equipartition field strength under some optimistic assumptions ($k=1/2$), though typically it appears to be below the equipartition field strength. We overall conclude that the suppression of convection via the initial magnetic field strength may be hard to achieve via the typically expected $B$-$n$
relations, though occasionally it may be feasible if the initial scaling relation is closer to $k=1/2$ and the recoupling of the magnetic field occurs early on. Nonetheless as noted by \citet{Tayler1987}, some significant effects may occur even for weaker fields, as the turbulent motions can be expected to tangle the field and reduce its scale, while simultaneously increasing its field strength. Once the equipartition field strength is reached locally on a scale $d$, then further tangling will be prohibited close to the magnetic flux rope. The field strength within the rope was shown by \citet{Tayler1987} to be 
\begin{equation}
B_{\rm rope}\sim B_i \left( \frac{R_*}{d} \right)^2,
\end{equation}
where $B_i$ is the field strength at protostar formation. The largest value can be obtained when $d\sim l_m$. We saw above that within the approximations used here, $d\sim R_*$, though this may have to be revisited via more detailed models.

\subsection{Estimates of protostellar resistivity}

Our subsequent considerations concerning the decay of the magnetic field require an estimate of the physical resistivity. In principle the physical resistivity can be calculated as the Spitzer resistivity via
\begin{equation}
    \eta_{\rm S} \sim \frac{Z_i e^2 m_e^{1/2} \ln \Lambda}{(k_{\rm B} T)^{3/2}} \sim 300 \left( \frac{T}{5 \times 10^6 ~\mathrm{K}} \right)^{-3/2}~\mathrm{cm^2\, s^{-1}}
\end{equation}
with $Z_i=1$ and $\ln\Lambda\sim20$ the Coulomb logarithm. The main dependence is then on the temperature $T$, which can be evaluated from the protostellar model. 
In a turbulent system, the effective resistivity is however more likely to be given by turbulent resistivity. Following \citet{vaisala14}, we have
\begin{equation}
    \eta_{\rm turb} = \frac{\varv_c}{3 k_c}
\end{equation}
with $k_c = 2\pi/l_m$, where we have taken the driving scale of the convection $l_m$ to be comparable to the  mixing length. Following this model, a summary of our estimated Spitzer and turbulent resistivity parameters is given in Table~\ref{tabla2}. We overall conclude that physical resistivity parameters should be roughly in the range of $10^{13}-10^{15}$~cm$^{2}$~s$^{-1}$.

\begin{table}[t!]
\centering
\caption{Estimations of the Spitzer and turbulent resistivity parameters for the fully convective protostellar models adopted by \citet{Moss2003}.}
\begin{tabular}{lcccc}
\hline\hline\noalign{\smallskip}
$R_*$ & $T_*$  & $\varv_c$ & $\eta_{\rm S}$ & $\eta_{\rm turb}$
\\
$(R_\odot)$ & $(\mathrm{K})$  & $(\mathrm{m~s^{-1}})$ & $(\mathrm{cm^2~s^{-1}})$ & $(\mathrm{cm^2~s^{-1}})$
\\
\hline
$M_*=1~M_\odot$ \\
\hline
13 & 1.1$\times 10^6$ & 37.94 & 3.1$\times 10^3$ & 1.8$\times 10^{14}$  \\
6 & 2.3$\times 10^6$ & 33.35 & 9.6$\times 10^2$ & 7.4$\times 10^{13}$  \\
3 & 4.6$\times 10^6$ & 29.71 & 3.4$\times 10^2$ & 3.3$\times 10^{13}$  \\
2 & 6.9$\times 10^6$ & 27.77 & 1.9$\times 10^2$ & 2.1$\times 10^{13}$  \\
1.8 & 7.7$\times 10^6$ & 27.29 & 1.6$\times 10^2$ & 1.8$\times 10^{13}$  \\
\hline
$M_* = 3~M_\odot$\\
\hline
22 & 1.9$\times 10^6$ & 215.22 & 1.3$\times 10^3$ & 1.7$\times 10^{15}$  \\
15 & 2.8$\times 10^6$ & 201.91 & 7.3$\times 10^2$ & 1.1$\times 10^{15}$  \\
13 & 3.2$\times 10^6$ & 197.15 & 5.9$\times 10^2$ & 9.5$\times 10^{14}$  \\
11 & 3.8$\times 10^6$ & 191.74 & 4.6$\times 10^2$ & 7.8$\times 10^{14}$  \\
9 & 4.6$\times 10^6$ & 185.43 & 3.4$\times 10^2$ & 6.2$\times 10^{14}$  \\
\hline
$M_* = 5~M_\odot$ \\
\hline
45 & 1.5$\times 10^6$ & 521.73 & 1.8$\times 10^3$ & 8.7$\times 10^{15}$  \\
30.5 & 2.3$\times 10^6$ & 488.98 & 9.9$\times 10^2$ & 5.5$\times 10^{15}$  \\
21.6 & 3.2$\times 10^6$ & 461.66 & 5.9$\times 10^2$ & 3.7$\times 10^{15}$  \\
19.3 & 3.6$\times 10^6$ & 453.07 & 5.0$\times 10^2$ & 3.2$\times 10^{15}$  \\
18 & 3.8$\times 10^6$ & 447.84 & 4.5$\times 10^2$ & 3.0$\times 10^{15}$ \\
\hline
\end{tabular}
\label{tabla2}
\end{table}

\subsection{Field decay due to convection}\label{decay}
In the following, we consider the evolution of the magnetic field in a fully convective protostar. First, in the case of flux freezing, adopting an initial radius $R_i$ of the protostar and an initial magnetic field $B_i$, this implies
\begin{equation}
B=\left( \frac{R_i}{R_*}\right)^2B_i.
\end{equation}
In the case of turbulent decay, we consider the induction equation,
\begin{equation}
\frac{d\vec{B}}{dt}=-\vec{\nabla}\times(\eta_{\rm turb} \vec{\nabla}\times\vec{B}),
\end{equation}
where $d/dt$ denotes the Lagrangian derivative following the radial inflow as the star contracts. If  $l_B$ is the characteristic length scale of the magnetic field, we can approximately write this diffusion equation as
\begin{equation}
\frac{dB}{dt}\sim - \frac{\eta_{\rm turb}}{l_B^2} B.
\end{equation}
The resulting solution corresponds to an exponential decay. If we also consider the effect of contraction of the protostar, an approximate evolution equation is given as
\begin{equation}
B(t)=B_i \left( \frac{R_i}{R_*} \right)^2\mathrm{\ exp}\left( -\frac{\eta_{\rm turb}}{l_B^2} t \right).
\end{equation}
We note that the equation is based on some simplifying assumptions and in particular may not capture some effects due to the inhomogeneity of the magnetic field. Particularly if the field strength is locally above the equipartition  value, our model may overestimate the amount of turbulent decay. It however serves to provide an approximate upper limit concerning the amount of decay that could occur.

\begin{figure}
    \centering
    \includegraphics[scale=0.55]{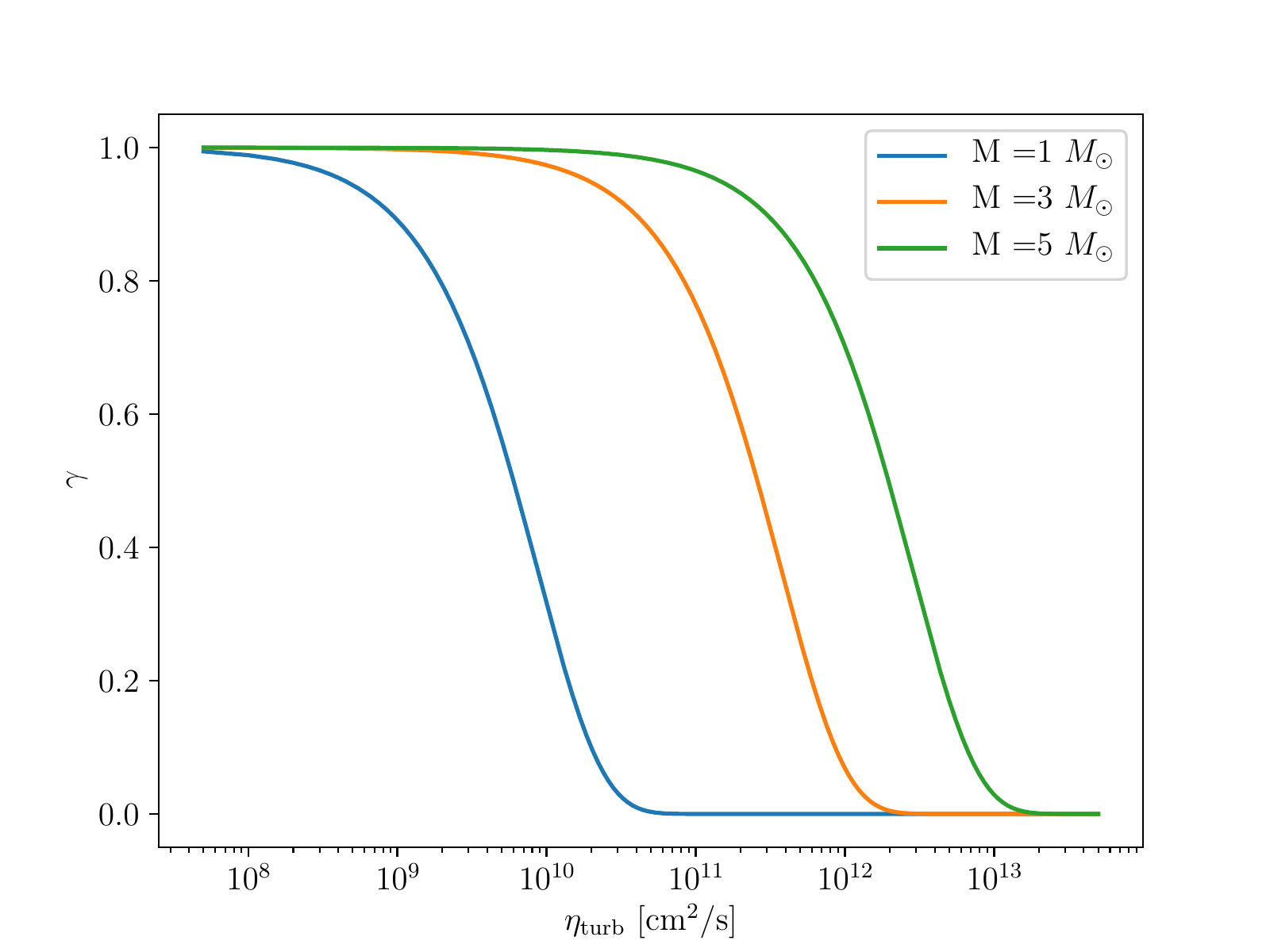}
    \caption{The fraction of magnetic flux that survives the turbulent decay during the evolution of a fully convective protostar as a function of the turbulent resistivity $\eta_{\rm turb}$.}
    \label{fig:decay}
\end{figure}

In Fig.~\ref{fig:decay}, the fraction of the magnetic flux that survives the decay of the magnetic field is given as a function of the  turbulent resistivity $\eta_{\rm turb}$ for different protostellar models. Our results are similar to \citet{Moss2003} in that we find more massive protostellar models to favor the survival of magnetic flux compared to lower-mass ones. However, it is important to note that around  turbulent resistivities of $\eta_{\rm turb}\sim10^{12}-10^{13}$~cm$^2$~s$^{-1}$, the surviving flux decreases considerably and approaches zero. A comparison with Table~\ref{tabla2}) however shows that the expected turbulent resistivities are always larger by more than two orders of magnitude than what is needed for significant fractions of the magnetic field to survive.

{It is well known from observations that fully convective stars can have strong dynamo-generated magnetic fields. This has been studied in great detail for the long period system GJ 65 \citep{Kochukhov2017, Shulyak2017}, where both of the fully convective M-dwarf components have masses of approximately $0.12$~M$_\odot$ \citep{Benedict2016,Kervella2016}, and similarly also for other low mass M-dwarf systems \citep{Morin2010}. For such a dynamo to operate, the dynamo number $D$ defined by Eq.~(\ref{dynamonumber}) needs to be larger than $7$. Assuming fixed convective velocities in the protostars as given in Table~\ref{tabla2}, the latter requirement essentially translates into a condition on the angular velocity of the star, given as}
\begin{equation}
\Omega>\sqrt{\frac{7}{9}}\frac{v_c}{R}.
\end{equation}
{For a typical convective velocity $v_c\sim30$~m~s$^{-1}$ from Table~\ref{tabla2}, and a stellar radius of about $2R_\odot$, this condition translates into a rotation period of about $97$~days or less. The operation of a dynamo in principle thus can be expected to be common. However, as we discuss in Section~\ref{conclusions} and as argued by \citet{Braithwaite2017}, such a dynamo-generated field is likely to significantly decay when the star transitions towards a radiative phase, and the remaining large scale field would be determined via helicity conservation.
}

\section{Magnetic field evolution in protostars transitioning from a convective to a radiative state}\label{radiative}

Here in the first subsection, we will consider the accretion of magnetized material onto radiative protostars. In the second subsection, we consider the interplay of accretion and decay during the previous convective phase.

\subsection{Accretion onto radiative protostars}\label{accradiative}

Already \citet{Dudorov1990} suggested that the transition of a protostar from the convective to the radiative phase will be crucial for the survival of the magnetic field, and that a critical mass for this transition could explain why chemically peculiar stars seem to have at least about $2$~M$_\odot$. Indeed, in protostellar models  with constant accretion rates, \citet{Palla1991, Palla1992} found such a transition to occur. Assuming an accretion rate of $\dot{M_*}=10^{-5}$~M$_\odot$~yr$^{-1}$, they found this transition to happen when the protostar reaches $M_{\rm crit,5}=2.4$~M$_\odot$. In case of a larger accretion rate of $\dot{M_*}=10^{-4}$~M$_\odot$~yr$^{-1}$, the transition occurred later at $M_{\rm crit,4}=4.5$~M$_\odot$ \citep{Palla1992}. 

\begin{figure}
    \centering
    \includegraphics[scale=0.55]{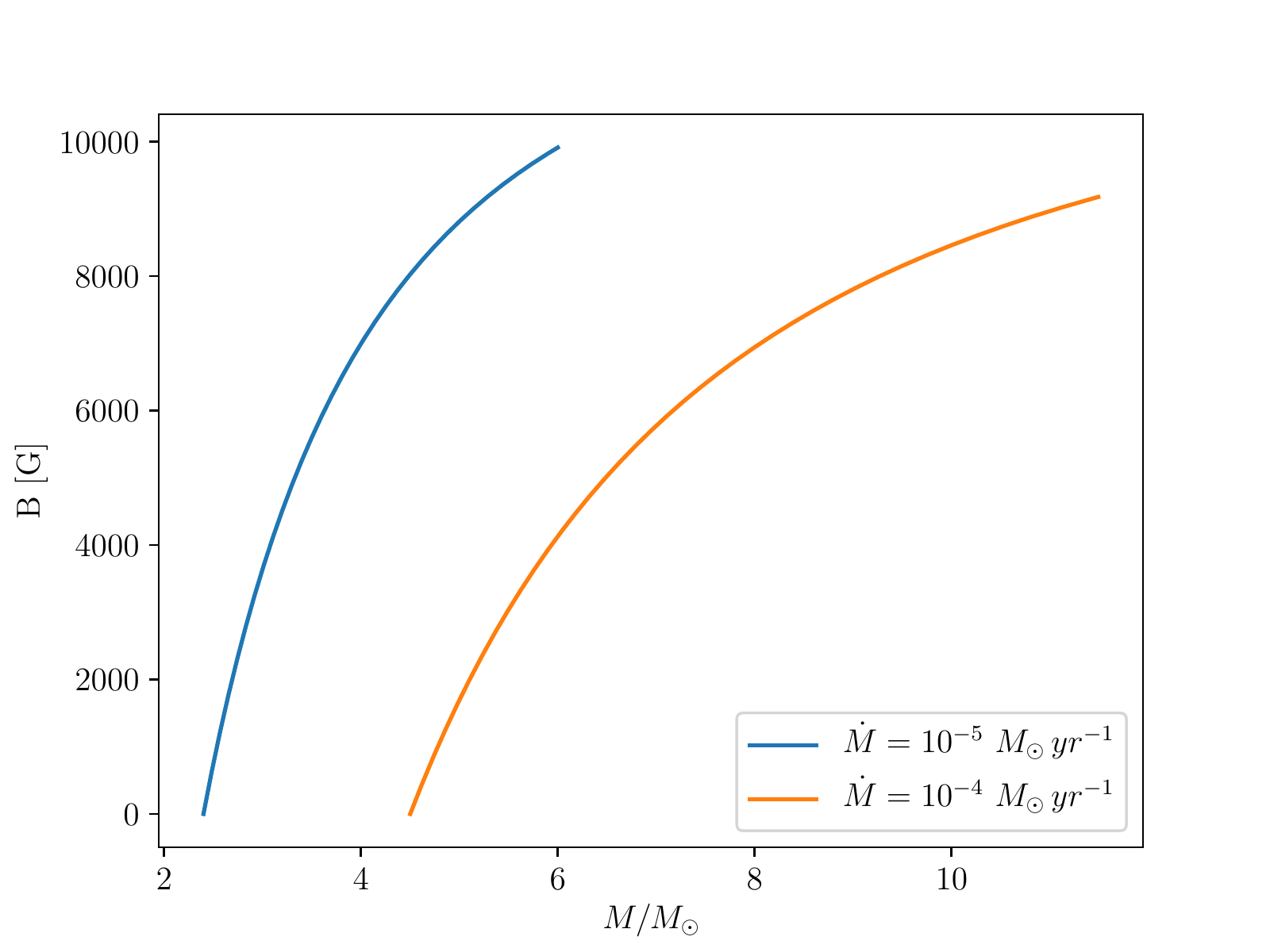}
    \caption{Expected time evolution of the fossil field in the star considering the models by \citet{Palla1991, Palla1992} with a transition from a fully convective to a radiative phase, for accretion rates of ${\dot M}_*=10^{-5}$~M$_\odot$~yr$^{-1}$ and $10^{-4}$~M$_\odot$~yr$^{-1}$.}

    \label{fig:palla}
\end{figure}

The transport of magnetic flux onto the protostar in general is a difficult problem. We consider here the dimensionless mass-to-flux ratio \citep{Nakano1978,Basu1994} in the form\begin{equation}
\lambda=\frac{2\pi G^{1/2}M}{\Phi},
\end{equation}
where $M$ denotes the associated mass within radius $R$, $\Phi\sim \pi R^2 B$ is the magnetic flux and $B$ the mean field strength penetrating through the projected surface of the volume that is considered. We recall that the critical mass-to-flux ratio is given by
\begin{equation}
\left(\frac{M}{\Phi}\right)_{\rm crit}=\frac{1}{2\pi G^{1/2}},
\end{equation}
and corresponds to $\lambda=1$.
For gravitational instability to occur, an initial condition of 
$\lambda \gtrsim 1$ is needed. Assuming ideal magneto-
hydrodynamics, \citet{Galli2006} and \citet{Shu2007} derived a mass-to-flux ratio $\lambda_{\rm d}\sim4$ for a newly-formed 
protostellar disk. On the other hand, observations of T-Tauri stars have inferred mass-to-flux ratios $\lambda_* \sim 10^3$--$10^4$ \citep{Krull2004}. In chemically peculiar stars, the most strongly magnetized stars have mass-to-flux ratios $\lambda_* \sim 10^3$, while more typical values are around $10^8$ \citep{Braithwaite2017}. 

In this regard, we recall our discussion in Section~\ref{prestellar} based on the disk model of \citet{Dudorov2014}, who find extremely low ionization degrees between scales of $0.2$ and $10$~au, such that the magnetic field decouples from the gas and ambipolar diffusion becomes relevant. As a result, they find magnetic field strength on both scales that are approximately the same, implying a decrease of the magnetic flux by a factor $\alpha_\lambda= (0.2/10)^2\sim 4\times10^{-4}$. 

To derive an evolution model for the magnetic flux $\Phi_*$ within the protostar, we will assume in the following that any initial magnetic field during the convective phase will decay very efficiently, until the critical mass $M_{\rm crit}$ is reached when the protostar becomes radiative. We then assume it to evolve according to the following evolution equation,
\begin{equation}
\dot{\Phi}_*=\alpha_\lambda \left[\lambda_{\rm d} \left(\frac{M}{\Phi}\right)_{\rm crit}\right]^{-1} \dot{M_*},\label{phidot}
\end{equation}
where 
$\lambda_{\rm d}\sim4$ corresponds to the mass-to-flux ratio of the disk normalized by the critical mass-to-flux ratio $(M/\Phi)_{\rm crit}$, and $\alpha_\lambda\sim4\times10^{-4}$ parametrises the magnetic flux loss during the disk phase.

In order to calculate the typical stellar magnetic field strength from the magnetic flux, we consider the protostellar models by \citet{Palla1991} and \citet{Palla1992}, with accretion rates $\dot{M_*}=10^{-5}$~M$_\odot$~yr$^{-1}$ and $10^{-4}$~M$_\odot$~yr$^{-1}$, respectively, allowing to estimate the mean magnetic field strength as $B_*\sim \Phi_*/(\pi R_*^2)$, with $R_*$ denoting the protostellar radius. The quantity of interest from an observational point of view is  the magnetic field strength when the star has contracted to the main sequence, for which we adopt an approximate mass-radius relation \begin{equation}
\left( \frac{R_*}{R_\odot} \right)\sim \left( \frac{M_*}{M_\odot}\right)^{0.57}.
\end{equation}
Evaluating the magnetic field strength with this quantity and plotting it as a function of $M_*$, the only difference between the cases with different accretion rates are due to the effect of the different critical mass when the transition occurs. The results from the two specific models with $\dot{M_*}=10^{-5}$~M$_\odot$~yr$^{-1}$ and $10^{-4}$~M$_\odot$~yr$^{-1}$ are given in Fig.~\ref{fig:palla}. We of course note that even more variability is conceivable, considering that the accretion rate could have different values and also be dependent on time.

\subsection{Turbulent decay in the presence of accretion}

In the calculation above, as we considered the radiative protostellar phase, we assumed that magnetic flux is conserved once it is accreted onto the protostar. During the previous convective phase, on the other hand, it is exposed to turbulent decay. While in Section~\ref{accradiative} we made the simplifying assumption that this decay was very efficient, with no relevant magnetic field being left over from the convective phase, we now relax this assumption and consider the interplay of accretion and decay. For this purpose, we will formulate Eq.~(\ref{phidot}), the evolution equation of the protostellar magnetic flux, as an evolution equation for the magnetic field. This further allows to combine it with the framework to describe the decay of the magnetic field developed in Section 3. We obtain:
\begin{equation}
\dot{B}_*=\alpha \left[\pi R_*^2\lambda_{\rm d} \left(\frac{M}{\Phi}\right)_{\rm crit}\right]^{-1} \dot{M_*}-\eta_{\rm turb} \frac{B_*}{l_B^2}\label{Bdot},
\end{equation}
where $R_*$ is the protostellar radius when the protostar reaches a mass $M_*$. We recall that $\eta$ denotes the turbulent resistivity introduced above, while $l_B$ is the characteristic length scale of the magnetic field. With the magnetic field being determined by the protostellar accretion process, we adopt $l_B\sim R_*$.

\begin{figure}
    \centering
    \includegraphics[scale=0.55]{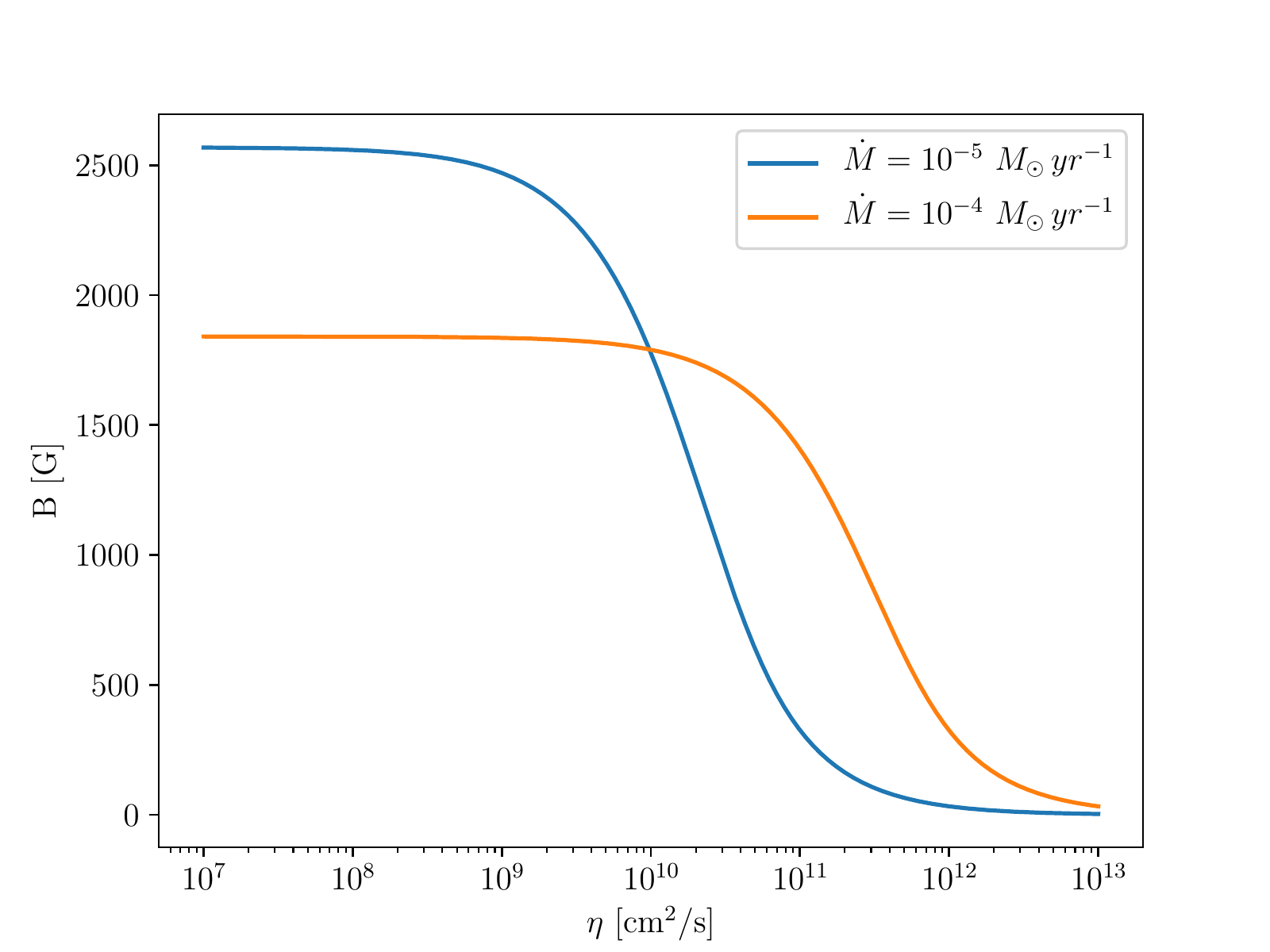}
    \caption{Expected fossil field as a function of $\eta_{\rm turb}$, considering turbulent decay in the presence of accretion at the time $t_f = M_\mathrm{crit}/\dot{M}$. We adopt here the characteristic properties of the \citet{Palla1991, Palla1992} models for the convective phase of the protostar.}
    \label{fig:acc-decay}
\end{figure}

Following \citet{Palla1991}, we consider the case with a protostellar accretion rate $\dot{M_*}=10^{-5}$~M$_\odot$~yr$^{-1}$, adopting an approximate protostellar radius of $10^{11.5}$~cm. The protostar remains convective until it reaches a mass of $2.5$~M$_\odot$. Similarly, considering \citet{Palla1992}, 
we adopt a protostellar radius of $10^{11.7}$~cm. It remains convective until reaching a mass of $4.5$~M$_\odot$. The strength of the magnetic
field at the time $t_f=M_{\rm crit}/{\dot M_*}$ when the mass of the accreting protostar reaches $M_{\rm crit}$, is 
shown in Fig.~\ref{fig:acc-decay} as a function of $\eta_{\rm turb}$. Our results are similar to the results for the $5$~M$_\odot$ protostar considered in Section~\ref{decay}, though still imply that the magnetic field would decay efficiently in the presence of realistic resistivity parameters.

\section{Protostellar models with a radiative core}\label{core}

Protostellar evolution is complex, and the behaviour in the outer layers may depend sensitively on how accretion happens as a function of time. In the simulations by \citet{Klassen2012}, the evolution and accretion of a massive protostar was followed, developing a radiative core after some time. We here explore the implications of such a model. We assume the radiative core to form at a critical mass $M_{\rm crit}$, and denote the mass of the core as $M_{\rm core}$ and its radius as $R_{\rm core}$. For simplicity we assume the core mass to be constant, while the mass of the envelope is expected to increase due to the accretion rate $\dot{M_*}$. We also assume the accretion rate to be constant, at least on average. 

Here we adapt our model for convection in a fully convective star to the case of a convective envelope around a radiative core. We have\begin{equation}
F_{\rm conv}\sim \frac{L_*}{4\pi R_*^2},\quad  \rho_{\rm env}\sim\frac{3(M_*-M_{\rm core})}{4\pi (R_*^3 -R_{\rm core}^3 )}\sim \frac{3(M_*-M_{\rm core})}{4\pi R_*^3}.
\end{equation}
From the equation of hydrostatic equilibrium,
\begin{equation}
\frac{dP}{dr}=-\frac{Gm}{r^2}\rho,
\end{equation}
with $P$ the pressure, $r$ the radial coordinate and $m$ the mass enclosed in radius $r$. The pressure in the envelope $P_{\rm env}$ can be estimated via
\begin{equation}
\frac{P_{\rm env}}{R_*-R_{\rm core}}\sim \frac{GM_{\rm core}} {R_*^2}\rho_{\rm env}.
\end{equation}
The relation between pressure $P_{\rm env}$ and temperature $T_{\rm env}$ is given as
\begin{equation}
P_{\rm env}=\frac{\rho_{\rm env}}{\mu m_p}k_{\rm B} T_{\rm env}.
\end{equation}
We thus obtain
\begin{equation}
T_{\rm env}=\frac{GM_{\rm core}(R_*-R_{\rm core})}{R_*^2}\frac{\mu m_p}{k_{\rm B}}.
\end{equation}
Using 
\begin{equation}
H_p=\left| \frac{dr}{d\ln P} \right| \sim \frac{P_{\rm env}}{\rho_{\rm env} g_{\rm env}},
\end{equation}
we further have:
\begin{eqnarray}
\sqrt{g_{\rm env} H_p} &=& \sqrt{\frac{k_{\rm B} T_{\rm env}}{\mu m_p}}\sim\sqrt{\frac{GM_{\rm core}(R_*-R_{\rm core})}{R_*^2}},\nonumber \\
 g_{\rm env}&\sim& \frac{GM_{\rm core}}{R_{\rm *}^2}.\label{envelope}
\end{eqnarray}
Inserting these expressions in Eq.~(\ref{Fconv}), we obtain\begin{eqnarray}
\nabla-\nabla_{\rm ad} & = & \left( \frac{2\sqrt{2}}{15} \right)^{2/3} \left( \frac{H_p}{l_m}  \right)^{4/3}
 \nonumber\\
&\times & \frac{L_*^{2/3}R_*^{8/3}}{G M_{\rm core}(M_*-M_{\rm core})^{2/3}(R_*-R_{\rm core})}.
\end{eqnarray}

Using this expression, we can calculated the convective velocity in the protostar using Eq.~(\ref{vc}), employing the speed of sound in the envelope, which we estimate as 
\begin{equation}
\varv_{\rm s,env}=\sqrt{\frac{GM_{\rm core}}{R_*}}.
\end{equation}
Using this formalism, we can follow the decay of the magnetic field as in Section~\ref{decay}. Particularly, we adopt here the models for the protostars with $3$~M$_\odot$ and $5$~M$_\odot$, but modify them to include a radiative core, which we take to be $1$~M$_\odot$ and $2.5$~M$_\odot$, respectively. We take the size of the radiative core to be $30\%$ of the protostellar radius. Under these assumptions, we calculate the evolution of the magnetic field within the radiative core. The results are included in Fig.~\ref{fig:bn} for comparison with the fully convective models. We find that the required field strength to suppress convection are very similar but slightly reduced compared to the case of fully convective protostars. In principle we thus conclude that a radiative core would not strongly alter the evolution of the magnetic field.

\section{Interaction of fossil field with core dynamo}\label{dynamo}

We have seen in the above that fossil fields can survive the protostellar phase if the protostar becomes radiative so that the field is not exposed to turbulent decay, or if convection in the protostar is suppressed or unexpectedly inefficient. An alternative possibility that is sometimes discussed concerns the origin of magnetic fields in the convective core of radiative stars, or the possible interaction of a core dynamo with the fossil field. Indeed, the cores of A-type stars indeed have long been suspected to harbor dynamo action \citep{Krause1976}, a possibility that has been confirmed via 3D magneto-hydrodynamical simulations by \citet{Brun2005}. The presence of such a dynamo may produce magnetically buoyant structures which may rise to the surface, a possibility that was confirmed by \citet{MacGregor2003} via numerical simulations for O and B stars. On the other hand, \citet{MacDonald2004} have pointed out that including realistic compositional gradients would severely prohibit the rise of magnetic flux tubes, and requiring a very high field strength for this mechanism to work, for plasma $\beta$ parameters (defined as the ratio of thermal over magnetic pressure) of the order $0.1$, which may then produce surface magnetic field strengths in the range of $\sim100-1690$~G in their models. Obtaining such low values of plasma $\beta$ may however be a significant challenge, and the small flux tube required for the rise of the magnetic field are not necessarily compatible with the coherent magnetic field structures suggested by observations.

\begin{figure}
    \centering
    \includegraphics[scale=0.18]{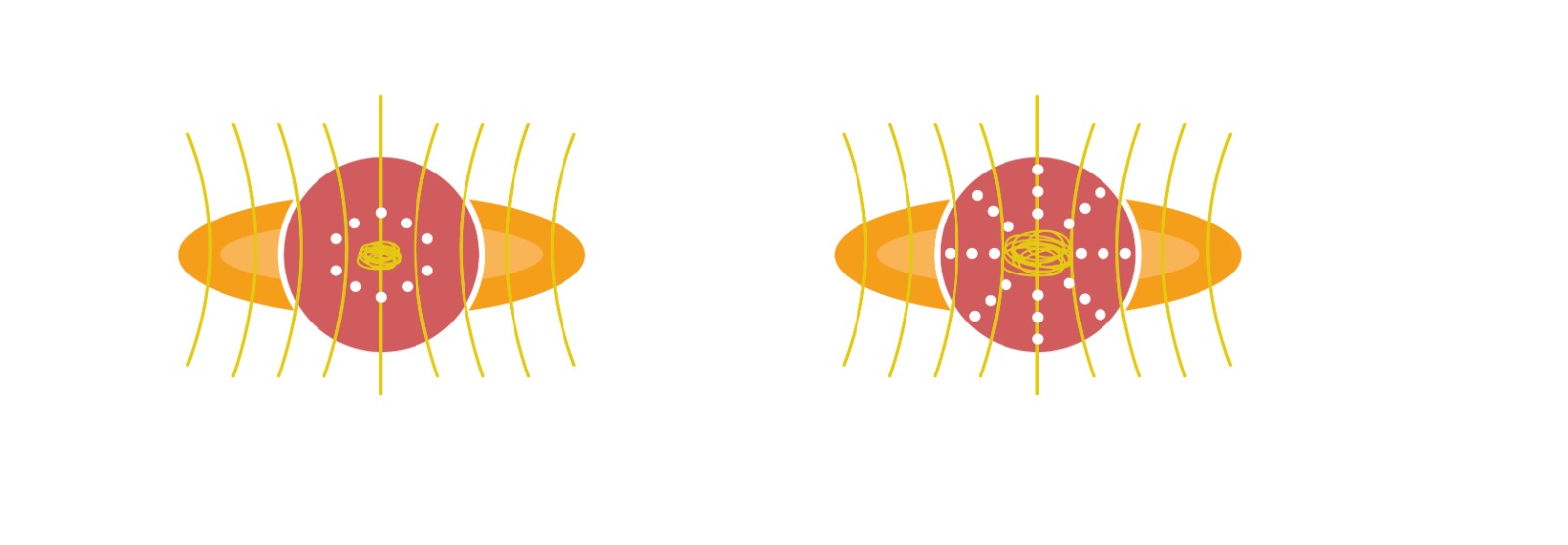}
    \caption{Interaction of fossil field via core dynamo. Left: Small convective core, amplification of magnetic field but not strong enough to rise to the surface. Right: Large massive core, production of strong fields, flux tubes reaching the stellar surface. {Circles indicate convective regions within the protostars, arrows represent radiative regions.}}
    \label{fig:dinamo}
\end{figure}

Normally in the presence of a core dynamo, one would expect field strength of the order to the equipartition field strength introduced above (Eq.~\ref{Bequi}). In the presence of a toroidal fossil field, however, \citet{Featherstone2009} have shown that the behavior of the dynamo changes, and it produces magnetic energies about $10$ times the kinetic energy of the convection, i.e. $B_{\rm core}\sim3B_{\rm eq}$. In the simulations by \citet{Featherstone2009}, the probability distribution function shows a power-law behavior scaling as $B^{-1}$, and the emergence of such types of power-laws is also suggested in the review by \citet{Brandenburg2005}. \citet{MacDonald2004} considered the rise of flux tubes with sizes between $10^{-2}$ and $10^{-4}$ times the pressure scale height, indicating that these structures are at least a few orders of magnitude smaller than the radius of the core. Assuming such a power-law type behavior, we then consider peak magnetic field strength of $B_{\rm peak}\sim\alpha B_{\rm core}\sim10\alpha B_{\rm eq}$, where as a fiducial value we adopt $\alpha\sim10^3$ that could be attained on the small scales considered here. The corresponding magnetic pressure is then given as 
\begin{equation}
P_{\rm mag}=\frac{B_{\rm peak}^2}{8\pi}.
\end{equation}
We estimate thermal pressure in the core  via 
\begin{equation}
P_{\rm therm}=\frac{\rho_{\rm core}}{\mu m_p}k_{\rm B} T_{\rm core},
\end{equation}
and $T_{\rm core}$ can be estimated via the expressions in Section~\ref{convection}. We here adopt the protostellar evolution models of \citet{Palla1991, Palla1992} with accretion rates of $10^{-5}$~M$_\odot$~yr$^{-1}$ and $10^{-4}$~M$_\odot$~yr$^{-1}$, respectively. In the first case, the convective core has a mass of about $1$~M$_\odot$ with a radius of $\sim 0.45$~R$_\odot$ (final stellar mass $8$~M$_\odot$ and radius $3.6$~R$_\odot$), in the second case the core consists of about $3$~M$_\odot$ with a radius of $\sim0.9$~R$_\odot$ (final stellar mass $15$~M$_\odot$ and radius $5.7$~R$_\odot$). The respective values of $\beta=P_{\rm therm}/P_{\rm mag}$ are given in Table~\ref{tabdynamo}; we of course emphasize that these are  estimates with some uncertainties and the mechanism requires the presence of a fossil field with toroidal structure to be operational. For the case with a convective core of $\sim1$~M$_\odot$, our estimated value of plasma $\beta$ is of the order $5$, and it seems unlikely to be able to fulfill the conditions where it could raise to the surface, as suggested by the criteria of \citet{MacDonald2004}. On the other hand, for the more massive core of $\sim3$~M$_\odot$ in the second model, we obtain $\beta\sim0.24$, and  a relevant contribution to the surface magnetic fields seems then conceivable. The two possible scenarios are summarized in Fig.~\ref{fig:dinamo}.

\begin{table}[t!]
\centering
\caption{Estimations of $\beta = P_\mathrm{therm}/P_{\rm mag}$ for flux tubes generated by the convective core dynamo due to the interaction with the fossil field. }
\begin{tabular}{lcccc}
\hline\hline\noalign{\smallskip}
$M_{\rm core}$ & $R_{\rm core}$  & $P_\mathrm{therm}$ & $ P_{\rm mag}$  & $\beta$ \\
$(M_\odot)$ & $(R_\odot)$  & (Pa) & (Pa) & \\
\hline
1 & 0.45 & $1.95\times 10^{14}$ & $3.6 \times 10^{14}$ & 5.38 \\
3 & 0.9 &  $1.10\times 10^{15}$ & $4.6\times 10^{15}$ & 0.237 \\
\hline
\end{tabular}
\label{tabdynamo}
\end{table}

\section{Discussion and conclusions}
\label{conclusions}

We have assessed here the possible origin of magnetic fields in chemically peculiar stars, considering the survival of interstellar medium magnetic fields during the pre-main sequence evolution of intermediate mass stars. For this purpose, we have considered a range of possible scenarios, including fully convective protostellar models as proposed by \citet{Siess2000}, for intermediate-mass protostellar models with a transition from a convective to a radiative phase \citep{Palla1991,Palla1992, Palla1993}, as well as models that include a radiative core during the protostellar evolution, as found e.g. by \citet{Klassen2012}. 

In the case of fully convective models, we find similar results as \citet{Moss2003}, in the sense that more massive protostars tend to somewhat favor the survival of magnetic flux within the star, though the expected turbulent resistivities in this regime are likely too strong so that efficient decay should be expected. Considering further the observed $B$-$n$ 
relation in the interstellar medium, as well as the expected flux loss due to ambipolar diffusion \citep[see, e.g.,][]{Desch2001,Nakano2002, Masson2016}, it seems difficult that the expected magnetic field strength would be sufficient to suppress the convective decay; such a scenario would require a steeper magnetic field strength - density relation, perhaps in occasional cases of a more spherical collapse, and/or an environment where ambipolar diffusion is less relevant, for example due to locally increased cosmic ray ionization.

A promising alternative are the Palla \& Stahler models suggesting a transition from a convective to a radiative phase in the protostar. Such a transition was already envisioned by \citet{Dudorov1990} to strongly alleviate the problem of the survival of magnetic fields, and our results show that such models may naturally explain the observed magnetic field strength in Ap/Bp stars. Such a mechanism may also naturally explain that the phenomenon of peculiar stars does not exist below the mass range of $1.5$~$M_\odot$, and could further explain the observations that the fraction of large-scale magnetic fields increases towards larger stellar masses \citep{Auriere2007}. To predict and compare theoretical expectations with observational results, more detailed studies are necessary both concerning the expected (time-dependent) accretion rates of the protostars, as well as their effect on the protostellar structure, to understand how the latter will affect the survival of the magnetic fields.

We further considered the possible interaction of magnetic flux accretion and turbulent decay, where we found the turbulent decay to be the dominant process and the presence of ongoing accretion to not significantly affect the scenario. Similarly, we checked about the implications of a radiative core in the protostellar model, though not finding it to significantly alter the overall picture. 

Regarding the presence of a core dynamo in radiative stars, in principle we find it difficult for the latter to significantly contribute to the observed structure. First, as shown by \citet{MacDonald2004}, small flux tubes with plasma $\beta$ parameters of order $0.1$ would be needed for them to rise to the surface and thus to potentially become observable. We find here that this may only be the case if there is an enhancement of the core dynamo due to the presence of a fossil field, as reported by \citet{Featherstone2009}, and if the core is sufficiently large, as predicted in the protostellar models with larger accretion rates by \citet{Palla1992}.

A possibility we did not pursue here in detail concerns the presence of a protostellar dynamo during the convective phase of the protostellar evolution {and the question of how much of such a dynamo-generated field could be left over when the convective zone retreats to the surface, as outlined e.g. by \citet{Stahler2004}. An explanation of the fields in chemically peculiar stars via such a scenario in principle is difficult. A qualitative description of this transition has been provided by \citet{Braithwaite2017} in their Section~4.5, on which we aim to follow up here.} The magnetic structures created by the dynamo are likely to be relatively thin, and may then at least partially annihilate each other, rather than creating a strong large-scale field. Due to being a conserved quantity, magnetic helicity is likely to be important in this process. Assuming that the magnetic fields produced by the dynamo have a length scale $l\sim\alpha R_*$, then the number of such volumes within the star corresponds to $R_*^3/l^3$. Assuming such a volume has a magnetic energy $E_B$, the total magnetic energy is $\alpha^{-3}E_B$. 

On the other hand, if we calculate the magnetic helicity, such a volume may have a helicity $E_B l$, though they will be randomly oriented. As a result, the total helicity can be estimate via a random walk, implying\begin{equation}
H_{\rm tot}=\alpha^{-1.5}E_B l=\alpha^{-0.5}E_BR_*.
\end{equation}
Now assuming that the field forms a configuration aligned over a stellar diameter, helicity conservation implies that the energy in that field would be $0.5\alpha^{-0.5}E_B$, i.e. a small fraction of the total magnetic energy produced, namely $0.5\alpha^{-2.5}$. Even with $\alpha\sim0.1$, this implies that a fraction of $\sim1/632$ of the total magnetic energy could go into the large-scale field, and likely the fraction could even be smaller in realistic scenarios. As further mentioned by \citet{Braithwaite2017}, it is not enough to explain some magnetic field strength that is observed, but more importantly an explanation is required why only sometimes strong magnetic fields are being produced, and models must be consistent with the observed bimodality in the stellar field strengths. This is not to say that such a dynamo effect would not be relevant, though; particularly the detection of weak magnetic fields in Sirius and Vega suggests that some minimum magnetic field strength might always be present, and a protostellar dynamo might be a mechanism that could guarantee that at least for a large range of cases.

\section*{Acknowledgement}
DRGS and JPH gratefully acknowledge support by the ANID BASAL projects ACE210002 and FB21003, as well as via Fondecyt Regular (project code 1201280). DRGS thanks for funding via the  Alexander von Humboldt - Foundation, Bonn, Germany.


\begin{thebibliography}{78}
\expandafter\ifx\csname natexlab\endcsname\relax\def\natexlab#1{#1}\fi

\bibitem[{{Abt} \& {Morrell}(1995)}]{Abt1995}
{Abt}, H.~A. \& {Morrell}, N.~I. 1995, \apjs, 99, 135

\bibitem[{{Abt} \& {Snowden}(1973)}]{Abt1973}
{Abt}, H.~A. \& {Snowden}, M.~S. 1973, \apjs, 25, 137

\bibitem[{{Auri{\`e}re} {et~al.}(2007){Auri{\`e}re}, {Wade}, {Silvester},
  {Ligni{\`e}res}, {Bagnulo}, {Bale}, {Dintrans}, {Donati}, {Folsom},
  {Gruberbauer}, {Hui Bon Hoa}, {Jeffers}, {Johnson}, {Landstreet},
  {L{\`e}bre}, {Lueftinger}, {Marsden}, {Mouillet}, {Naseri}, {Paletou},
  {Petit}, {Power}, {Rincon}, {Strasser}, \& {Toqu{\'e}}}]{Auriere2007}
{Auri{\`e}re}, M., {Wade}, G.~A., {Silvester}, J., {et~al.} 2007, \aap, 475,
  1053

\bibitem[{{Basu} \& {Mouschovias}(1994)}]{Basu1994}
{Basu}, S. \& {Mouschovias}, T.~C. 1994, \apj, 432, 720

\bibitem[{{Benedict} {et~al.}(2016){Benedict}, {Henry}, {Franz}, {McArthur},
  {Wasserman}, {Jao}, {Cargile}, {Dieterich}, {Bradley}, {Nelan}, \&
  {Whipple}}]{Benedict2016}
{Benedict}, G.~F., {Henry}, T.~J., {Franz}, O.~G., {et~al.} 2016, \aj, 152, 141

\bibitem[{{Braithwaite}(2006)}]{Braithwaite2006}
{Braithwaite}, J. 2006, \aap, 453, 687

\bibitem[{{Braithwaite}(2009)}]{Braithwaite2009}
{Braithwaite}, J. 2009, \mnras, 397, 763

\bibitem[{{Braithwaite} \& {Spruit}(2004)}]{Braithwaite04}
{Braithwaite}, J. \& {Spruit}, H.~C. 2004, \nat, 431, 819

\bibitem[{{Braithwaite} \& {Spruit}(2017)}]{Braithwaite2017}
{Braithwaite}, J. \& {Spruit}, H.~C. 2017, Royal Society Open Science, 4,
  160271

\bibitem[{{Brandenburg} \& {Subramanian}(2005)}]{Brandenburg2005}
{Brandenburg}, A. \& {Subramanian}, K. 2005, \physrep, 417, 1

\bibitem[{{Brun} {et~al.}(2005){Brun}, {Browning}, \& {Toomre}}]{Brun2005}
{Brun}, A.~S., {Browning}, M.~K., \& {Toomre}, J. 2005, \apj, 629, 461

\bibitem[{{Carrier} {et~al.}(2002){Carrier}, {North}, {Udry}, \&
  {Babel}}]{Carrier2002}
{Carrier}, F., {North}, P., {Udry}, S., \& {Babel}, J. 2002, \aap, 394, 151

\bibitem[{{Chapman} {et~al.}(2013){Chapman}, {Davidson}, {Goldsmith}, {Houde},
  {Kwon}, {Li}, {Looney}, {Matthews}, {Matthews}, {Novak}, {Peng},
  {Vaillancourt}, \& {Volgenau}}]{Chapman2013}
{Chapman}, N.~L., {Davidson}, J.~A., {Goldsmith}, P.~F., {et~al.} 2013, \apj,
  770, 151

\bibitem[{{Cowling}(1945)}]{Cowling1945}
{Cowling}, T.~G. 1945, \mnras, 105, 166

\bibitem[{{Crutcher}(2012)}]{Crutcher12}
{Crutcher}, R.~M. 2012, \araa, 50, 29

\bibitem[{{Desch} \& {Mouschovias}(2001)}]{Desch2001}
{Desch}, S.~J. \& {Mouschovias}, T.~C. 2001, \apj, 550, 314

\bibitem[{{Dudorov} \& {Khaibrakhmanov}(2014)}]{Dudorov2014}
{Dudorov}, A.~E. \& {Khaibrakhmanov}, S.~A. 2014, \apss, 352, 103

\bibitem[{{Dudorov} \& {Khaibrakhmanov}(2015)}]{Dudorov2015}
{Dudorov}, A.~E. \& {Khaibrakhmanov}, S.~A. 2015, Advances in Space Research,
  55, 843

\bibitem[{{Dudorov} \& {Tutukov}(1990)}]{Dudorov1990}
{Dudorov}, A.~E. \& {Tutukov}, A.~V. 1990, \sovast, 34, 171

\bibitem[{{Featherstone} {et~al.}(2009){Featherstone}, {Browning}, {Brun}, \&
  {Toomre}}]{Featherstone2009}
{Featherstone}, N.~A., {Browning}, M.~K., {Brun}, A.~S., \& {Toomre}, J. 2009,
  \apj, 705, 1000

\bibitem[{{Folsom} {et~al.}(2013){Folsom}, {Wade}, \& {Johnson}}]{Folsom2013}
{Folsom}, C.~P., {Wade}, G.~A., \& {Johnson}, N.~M. 2013, \mnras, 433, 3336

\bibitem[{{Galli} {et~al.}(2006){Galli}, {Lizano}, {Shu}, \&
  {Allen}}]{Galli2006}
{Galli}, D., {Lizano}, S., {Shu}, F.~H., \& {Allen}, A. 2006, \apj, 647, 374

\bibitem[{{Gerbaldi} {et~al.}(1985){Gerbaldi}, {Floquet}, \&
  {Hauck}}]{Gerbaldi1985}
{Gerbaldi}, M., {Floquet}, M., \& {Hauck}, B. 1985, \aap, 146, 341

\bibitem[{{Girart} {et~al.}(2006){Girart}, {Rao}, \& {Marrone}}]{Girart2006}
{Girart}, J.~M., {Rao}, R., \& {Marrone}, D.~P. 2006, Science, 313, 812

\bibitem[{{Gough} \& {Tayler}(1966)}]{Gough1966}
{Gough}, D.~O. \& {Tayler}, R.~J. 1966, \mnras, 133, 85

\bibitem[{{Hayashi} {et~al.}(1962){Hayashi}, {H{\={o}}shi}, \&
  {Sugimoto}}]{Hayashi1962}
{Hayashi}, C., {H{\={o}}shi}, R., \& {Sugimoto}, D. 1962, Progress of
  Theoretical Physics Supplement, 22, 1

\bibitem[{{Hennebelle} {et~al.}(2016){Hennebelle}, {Commer{\c{c}}on},
  {Chabrier}, \& {Marchand}}]{Hennebelle2016}
{Hennebelle}, P., {Commer{\c{c}}on}, B., {Chabrier}, G., \& {Marchand}, P.
  2016, \apjl, 830, L8

\bibitem[{{Henrichs} {et~al.}(2000){Henrichs}, {de Jong}, {Donati}, {Wade},
  {Babel}, {Shorlin}, {Verdugo}, {Talavera}, {Catala}, {Veen}, {Nichols}, \&
  {Kaper}}]{Henrichs2000}
{Henrichs}, H.~F., {de Jong}, J.~A., {Donati}, D.~F., {et~al.} 2000, in
  Magnetic Fields of Chemically Peculiar and Related Stars, ed. Y.~V.
  {Glagolevskij} \& I.~I. {Romanyuk}, 57--60

\bibitem[{{Jermyn} \& {Cantiello}(2021)}]{Jermyn2021}
{Jermyn}, A.~S. \& {Cantiello}, M. 2021, \apj, 923, 104

\bibitem[{{Johns-Krull} {et~al.}(2004){Johns-Krull}, {Valenti}, \&
  {Saar}}]{Krull2004}
{Johns-Krull}, C.~M., {Valenti}, J.~A., \& {Saar}, S.~H. 2004, \apj, 617, 1204

\bibitem[{{Kervella} {et~al.}(2016){Kervella}, {M{\'e}rand}, {Ledoux},
  {Demory}, \& {Le Bouquin}}]{Kervella2016}
{Kervella}, P., {M{\'e}rand}, A., {Ledoux}, C., {Demory}, B.~O., \& {Le
  Bouquin}, J.~B. 2016, \aap, 593, A127

\bibitem[{{Kippenhahn} {et~al.}(2013){Kippenhahn}, {Weigert}, \&
  {Weiss}}]{Kippenhahn2013}
{Kippenhahn}, R., {Weigert}, A., \& {Weiss}, A. 2013, {Stellar Structure and
  Evolution}

\bibitem[{{Klassen} {et~al.}(2012){Klassen}, {Pudritz}, \&
  {Peters}}]{Klassen2012}
{Klassen}, M., {Pudritz}, R.~E., \& {Peters}, T. 2012, \mnras, 421, 2861

\bibitem[{{Kochukhov} {et~al.}(2011){Kochukhov}, {Makaganiuk}, {Piskunov},
  {Jeffers}, {Johns-Krull}, {Keller}, {Rodenhuis}, {Snik}, {Stempels}, \&
  {Valenti}}]{Kochukhov2011}
{Kochukhov}, O., {Makaganiuk}, V., {Piskunov}, N., {et~al.} 2011, \aap, 534,
  L13

\bibitem[{{Kochukhov} {et~al.}(2017){Kochukhov}, {Silvester}, {Bailey},
  {Landstreet}, \& {Wade}}]{Kochukhov2017}
{Kochukhov}, O., {Silvester}, J., {Bailey}, J.~D., {Landstreet}, J.~D., \&
  {Wade}, G.~A. 2017, \aap, 605, A13

\bibitem[{{Koley} {et~al.}(2022){Koley}, {Roy}, {Momjian}, {Sarma}, \&
  {Datta}}]{Koley2022}
{Koley}, A., {Roy}, N., {Momjian}, E., {Sarma}, A.~P., \& {Datta}, A. 2022,
  \mnras, 516, L48

\bibitem[{{Krause} \& {Oetken}(1976)}]{Krause1976}
{Krause}, F. \& {Oetken}, L. 1976, in IAU Colloq. 32: Physics of Ap Stars, ed.
  W.~W. {Weiss}, H.~{Jenkner}, \& H.~J. {Wood}, 29

\bibitem[{{Lavail} {et~al.}(2017){Lavail}, {Kochukhov}, {Hussain}, {Alecian},
  {Herczeg}, \& {Johns-Krull}}]{Lavail2017}
{Lavail}, A., {Kochukhov}, O., {Hussain}, G.~A.~J., {et~al.} 2017, \aap, 608,
  A77

\bibitem[{{Li} {et~al.}(2009){Li}, {Dowell}, {Goodman}, {Hildebrand}, \&
  {Novak}}]{Li2009}
{Li}, H.-b., {Dowell}, C.~D., {Goodman}, A., {Hildebrand}, R., \& {Novak}, G.
  2009, \apj, 704, 891

\bibitem[{{Ligni{\`e}res} {et~al.}(2009){Ligni{\`e}res}, {Petit}, {B{\"o}hm},
  \& {Auri{\`e}re}}]{Lignieres2009}
{Ligni{\`e}res}, F., {Petit}, P., {B{\"o}hm}, T., \& {Auri{\`e}re}, M. 2009,
  \aap, 500, L41

\bibitem[{{MacDonald} \& {Mullan}(2004)}]{MacDonald2004}
{MacDonald}, J. \& {Mullan}, D.~J. 2004, \mnras, 348, 702

\bibitem[{{MacGregor} \& {Cassinelli}(2003)}]{MacGregor2003}
{MacGregor}, K.~B. \& {Cassinelli}, J.~P. 2003, \apj, 586, 480

\bibitem[{{Markey} \& {Tayler}(1973)}]{Markey1973}
{Markey}, P. \& {Tayler}, R.~J. 1973, \mnras, 163, 77

\bibitem[{{Markey} \& {Tayler}(1974)}]{Markey1974}
{Markey}, P. \& {Tayler}, R.~J. 1974, \mnras, 168, 505

\bibitem[{{Masson} {et~al.}(2016){Masson}, {Chabrier}, {Hennebelle}, {Vaytet},
  \& {Commer{\c{c}}on}}]{Masson2016}
{Masson}, J., {Chabrier}, G., {Hennebelle}, P., {Vaytet}, N., \&
  {Commer{\c{c}}on}, B. 2016, \aap, 587, A32

\bibitem[{{Mathys}(2008)}]{Mathys2008}
{Mathys}, G. 2008, Contributions of the Astronomical Observatory Skalnate
  Pleso, 38, 217

\bibitem[{{Michaud}(1970)}]{Michaud1970}
{Michaud}, G. 1970, \apj, 160, 641

\bibitem[{{Morin} {et~al.}(2010){Morin}, {Donati}, {Petit}, {Delfosse},
  {Forveille}, \& {Jardine}}]{Morin2010}
{Morin}, J., {Donati}, J.~F., {Petit}, P., {et~al.} 2010, \mnras, 407, 2269

\bibitem[{{Moss}(1989)}]{Moss1989}
{Moss}, D. 1989, \mnras, 236, 629

\bibitem[{{Moss}(2003)}]{Moss2003}
{Moss}, D. 2003, \aap, 403, 693

\bibitem[{{Moss} \& {Tayler}(1969)}]{Moss1969}
{Moss}, D.~L. \& {Tayler}, R.~J. 1969, \mnras, 145, 217

\bibitem[{{Nakano}(1988)}]{Nakano1988}
{Nakano}, T. 1988, in NATO Advanced Study Institute (ASI) Series C, Vol. 232,
  Galactic and Extragalactic Star Formation, ed. R.~E. {Pudritz} \& M.~{Fich},
  111

\bibitem[{{Nakano} \& {Nakamura}(1978)}]{Nakano1978}
{Nakano}, T. \& {Nakamura}, T. 1978, \pasj, 30, 671

\bibitem[{{Nakano} {et~al.}(2002){Nakano}, {Nishi}, \&
  {Umebayashi}}]{Nakano2002}
{Nakano}, T., {Nishi}, R., \& {Umebayashi}, T. 2002, \apj, 573, 199

\bibitem[{{Nakano} \& {Umebayashi}(1988)}]{NakanoUmebayashi1988}
{Nakano}, T. \& {Umebayashi}, T. 1988, Progress of Theoretical Physics
  Supplement, 96, 73

\bibitem[{{Palla} \& {Stahler}(1991)}]{Palla1991}
{Palla}, F. \& {Stahler}, S.~W. 1991, \apj, 375, 288

\bibitem[{{Palla} \& {Stahler}(1992)}]{Palla1992}
{Palla}, F. \& {Stahler}, S.~W. 1992, \apj, 392, 667

\bibitem[{{Palla} \& {Stahler}(1993)}]{Palla1993}
{Palla}, F. \& {Stahler}, S.~W. 1993, \apj, 418, 414

\bibitem[{{Parker}(1979)}]{Parker1979}
{Parker}, E.~N. 1979, \apss, 62, 135

\bibitem[{{Petit} {et~al.}(2022){Petit}, {B{\"o}hm}, {Folsom}, {Ligni{\`e}res},
  \& {Cang}}]{Petit2022}
{Petit}, P., {B{\"o}hm}, T., {Folsom}, C.~P., {Ligni{\`e}res}, F., \& {Cang},
  T. 2022, \aap, 666, A20

\bibitem[{{Petit} {et~al.}(2011){Petit}, {Ligni{\`e}res}, {Auri{\`e}re},
  {Wade}, {Alina}, {Ballot}, {B{\"o}hm}, {Jouve}, {Oza}, {Paletou}, \&
  {Th{\'e}ado}}]{Petit2011}
{Petit}, P., {Ligni{\`e}res}, F., {Auri{\`e}re}, M., {et~al.} 2011, \aap, 532,
  L13

\bibitem[{{Petit} {et~al.}(2010){Petit}, {Ligni{\`e}res}, {Wade},
  {Auri{\`e}re}, {B{\"o}hm}, {Bagnulo}, {Dintrans}, {Fumel}, {Grunhut},
  {Lanoux}, {Morgenthaler}, \& {Van Grootel}}]{Petit2010}
{Petit}, P., {Ligni{\`e}res}, F., {Wade}, G.~A., {et~al.} 2010, \aap, 523, A41

\bibitem[{{Petit} {et~al.}(2013){Petit}, {Owocki}, {Wade}, {Cohen},
  {Sundqvist}, {Gagn{\'e}}, {Ma{\'\i}z Apell{\'a}niz}, {Oksala}, {Bohlender},
  {Rivinius}, {Henrichs}, {Alecian}, {Townsend}, {ud-Doula}, \& {MiMeS
  Collaboration}}]{Petit2013}
{Petit}, V., {Owocki}, S.~P., {Wade}, G.~A., {et~al.} 2013, \mnras, 429, 398

\bibitem[{{Schleicher} \& {Mennickent}(2017)}]{Schleicher2017}
{Schleicher}, D. R.~G. \& {Mennickent}, R.~E. 2017, \aap, 602, A109

\bibitem[{{Shu} {et~al.}(2007){Shu}, {Galli}, {Lizano}, {Glassgold}, \&
  {Diamond}}]{Shu2007}
{Shu}, F.~H., {Galli}, D., {Lizano}, S., {Glassgold}, A.~E., \& {Diamond},
  P.~H. 2007, \apj, 665, 535

\bibitem[{{Shukurov}(2004)}]{Shukurov2004}
{Shukurov}, A. 2004, arXiv e-prints, astro

\bibitem[{{Shulyak} {et~al.}(2017){Shulyak}, {Reiners}, {Engeln}, {Malo},
  {Yadav}, {Morin}, \& {Kochukhov}}]{Shulyak2017}
{Shulyak}, D., {Reiners}, A., {Engeln}, A., {et~al.} 2017, Nature Astronomy, 1,
  0184

\bibitem[{{Siess} {et~al.}(2000){Siess}, {Dufour}, \& {Forestini}}]{Siess2000}
{Siess}, L., {Dufour}, E., \& {Forestini}, M. 2000, \aap, 358, 593

\bibitem[{{Spruit}(1999)}]{Spruit1999}
{Spruit}, H.~C. 1999, \aap, 349, 189

\bibitem[{{Stahler} \& {Palla}(2004)}]{Stahler2004}
{Stahler}, S.~W. \& {Palla}, F. 2004, {The Formation of Stars}

\bibitem[{{Tayler}(1987)}]{Tayler1987}
{Tayler}, R.~J. 1987, \mnras, 227, 553

\bibitem[{{Tomida} {et~al.}(2015){Tomida}, {Okuzumi}, \&
  {Machida}}]{Tomida2015}
{Tomida}, K., {Okuzumi}, S., \& {Machida}, M.~N. 2015, \apj, 801, 117

\bibitem[{{V{\"a}is{\"a}l{\"a}} {et~al.}(2014){V{\"a}is{\"a}l{\"a}},
  {Brandenburg}, {Mitra}, {K{\"a}pyl{\"a}}, \& {Mantere}}]{vaisala14}
{V{\"a}is{\"a}l{\"a}}, M.~S., {Brandenburg}, A., {Mitra}, D., {K{\"a}pyl{\"a}},
  P.~J., \& {Mantere}, M.~J. 2014, \aap, 567, A139

\bibitem[{{Vall{\'e}e}(1997)}]{Vallee1997}
{Vall{\'e}e}, J.~P. 1997, \fcp, 19, 1

\bibitem[{{Wade} {et~al.}(2013){Wade}, {Grunhut}, {Petit}, {Neiner}, {Alecian},
  {Landstreet}, \& {MiMeS Collaboration}}]{Wade2013}
{Wade}, G.~A., {Grunhut}, J., {Petit}, V., {et~al.} 2013, in Massive Stars:
  From alpha to Omega, 51

\bibitem[{{Wright}(1973)}]{Wright1973}
{Wright}, G.~A.~E. 1973, \mnras, 162, 339

\bibitem[{{Wurster} {et~al.}(2018){Wurster}, {Bate}, \& {Price}}]{Wurster2018}
{Wurster}, J., {Bate}, M.~R., \& {Price}, D.~J. 2018, \mnras, 481, 2450

\bibitem[{{Wurster} {et~al.}(2022){Wurster}, {Bate}, {Price}, \&
  {Bonnell}}]{Wurster2022}
{Wurster}, J., {Bate}, M.~R., {Price}, D.~J., \& {Bonnell}, I.~A. 2022, \mnras,
  511, 746

\end{thebibliography}


%

%
%
%

%
%

%
%
\end{document}